\documentclass[twocolumn]{aastex631}
\pdfoutput=1 %for arXiv submission

\begin{document}

\title{Multiple misaligned outflows and warped accretion flows in the proto-multiple system Per-emb-8 and 55}

\author[0009-0006-6233-3363]{Shang-Jing Lin}
\affiliation{Institute of Astronomy, National Tsing Hua University, No. 101, Section 2, Kuang-Fu Rd., Hsinchu 30013, Taiwan}
\affiliation{Academia Sinica Institute of Astronomy and Astrophysics,
11F of Astronomy-Mathematics Building, AS/NTU, No.1, Sec. 4, Roosevelt Rd, Taipei 10617, Taiwan}

\author[0000-0003-1412-893X]{Hsi-Wei Yen}
\affiliation{Academia Sinica Institute of Astronomy and Astrophysics,
11F of Astronomy-Mathematics Building, AS/NTU, No.1, Sec. 4, Roosevelt Rd, Taipei 10617, Taiwan}

\author[0000-0001-5522-486X]{Shih-Ping Lai}

\affiliation{Institute of Astronomy, National Tsing Hua University, No. 101, Section 2, Kuang-Fu Rd., Hsinchu 30013, Taiwan}
\affiliation{Center for Informatics and Computation in Astronomy, National Tsing Hua University, No. 101, Section 2, Kuang-Fu Rd., Hsinchu 30013, Taiwan}
\affiliation{Department of Physics, National Tsing Hua University, No. 101, Section 2, Kuang-Fu Rd., Hsinchu 30013, Taiwan}
\affiliation{Academia Sinica Institute of Astronomy and Astrophysics,
11F of Astronomy-Mathematics Building, AS/NTU, No.1, Sec. 4, Roosevelt Rd, Taipei 10617, Taiwan}



\begin{abstract}
To investigate the formation process of multiple systems, we have analyzed the ALMA archival data of the 1.3 mm continuum, $^{12}$CO (2-1) and C$^{18}$O (2-1) emission in a proto-multiple system consisting of a Class 0 protostar Per-emb-8 and a Class I protobinary Per-emb-55 $A$ and $B$. 
The 1.3 mm continuum emission is likely to primarily trace their protostellar disks, 
and the Keplerian disk rotation is observed in Per-emb-8 and Per-emb-55 $A$ in the emission lines.
In Per-emb-8, we identify two arm-like structures with a length of $\sim$ 1000 au connecting the eastern and western of its disk in the continuum and C$^{18}$O emission. Our analysis suggests that these arm-like structures are most likely infalling flows.
In the $^{12}$CO emission, we discover a second bipolar outflow associated with Per-emb-8. 
The two bipolar outflows in Per-emb-8 are possibly launched along the normal axes of the misaligned inner and outer parts of its warped protostellar disk.
In Per-emb-55, we find that the red- and blueshifted lobes of its bipolar outflow are misaligned by 90$^\circ$. 
The presence of the warped disk, multiple misaligned outflows, and asymmetric infalling flows suggest complex dynamics in proto-multiple systems, and these could be related to the tidal interactions between the companions and/or the turbulent environments forming this proto-multiple system.

\end{abstract}


\section{Introduction} \label{sec:Intro}

Multiple systems are the common outcome of the star formation process \citep{2010ApJS..190....1R, 2013ARA&A..51..269D}.
Theoretical simulations suggest that the formation of multiple systems can occur either by turbulent fragmentation in dense cores \citep{2004ApJ...600..769F,2010ApJ...725.1485O} or gravitational instabilities in disks \citep{2009MNRAS.400.1563S, 2010ApJ...708.1585K}. The former mechanism tends to form widely separated binaries \citep[$\gtrsim$500 au; ][]{2010ApJ...725.1485O} and the latter tends to form close binaries \citep[$\textless$100 au; ][]{2018ApJ...857...40S}. 
The mass accretion and outflow processes are expected to be more complicated and show complex structures in proto-multiple systems, compared to single systems. 
Theoretical simulations show that turbulent environments and tidal interaction between companions can result in multiple arm-like structures around protostars and can also cause the disks to warp and the outflows to precess \citep{2018MNRAS.475.5618B...Bate, 2015MNRAS.449L.123M,2024MNRAS.52710131T}.

These various structures have been observed in several binary and multiple systems at (sub-)millimeter \citep[e.g.,][]{2023MNRAS.522.2384M, 2023ApJ...953...82L, 2016Natur.538..483T, 2019ApJ...870...81T, 2014ApJ...789L...4T}, and optical/IR wavelengths \citep[e.g.,][]{2007AJ....133.2799A,2021ApJ...919...23E,2002ApJ...568..733M}.
Besides, recent high-resolution and high-sensitivity Atacama Large Millimeter/submillimeter Array (ALMA) observations have revealed asymmetric protostellar envelopes and warped accretion flows around several protostars \citep{2014ApJ...793....1Y,2020NatAs...4.1158P,2020ApJ...893...51S,2021A&A...653A.166C,2022ApJ...925...32T}.
These features are different from the conventional picture of axisymmetric protostellar envelopes \citep{1987ARA&A..25...23S} and may affect our understanding of mass accretion during the star formation process. 
However, the number of observational studies characterizing these physical processes is still limited. 
Studying multiple protostellar systems in the early stages of star formation can shed light on the mass accretion process under the influence of turbulent environments and tidal interaction. 

Two major surveys toward the Perseus molecular cloud have yielded comprehensive datasets for studying the multiplicity. 
The Karl G. Jansky Very Large Array (VLA) Nascent Disk and Multiplicity survey (VANDAM) at a resolution of 0$\farcs$065 (15 au) measured the multiplicity of 94 protostars (82 Class 0/I and 12 Class II)  \citep{2016ApJ...818...73T}. 
The Mass Assembly of Stellar Systems and Their Evolution with the Submillimeter Array (SMA) survey (MASSES) at resolutions of $\sim$1\arcsec--4${\arcsec}$ (300--1200 au) revealed the gas kinematics and distributions around 74 Class 0/I protostars in the continuum and lines at 1.3 and 0.9 mm \citep{2018ApJS..237...22S, 2019ApJS..245...21S}. 
These surveys provide a comprehensive sample in the Perseus molecular cloud to search for candidate proto-multiple systems showing a common envelope with complex structures, which may hint at tidal interaction or complex dynamics. 

From these survey results, we select a young triple system, Class 0 protostar Per-emb-8 and Class I protobinary Per-emb-55, to study the gas dynamics and star formation process within proto-multiple systems.
The distance to this triple system is $\sim$300 pc \citep{2018ApJ...869...83Z}.
The projected separation between Per-emb-8 and 55 is 9\farcs3 (2782 au), and that between source $A$ and $B$ in the protobinary Per-emb-55 is 0\farcs6 (185 au) \citep[][]{2022ApJ...925...39T}. 
The bolometric luminosity ($L_{bol}$) and temperature ($T_{bol}$) of Per-emb-8 are $2.60\pm 0.5L_{\odot}$ and $43.0\pm6.0$ K, and those of Per-emb-55 are $1.80\pm 0.8L_{\odot}$ and $309.0\pm64.0$ K, respectively \citep{2016ApJ...818...73T}. 

The SMA observation in the $^{12}$CO lines reveal that Per-emb-8 is associated with a north-south bipolar outflow with the blueshifted lobe in the north at a position angle (PA) of $\sim$15$^{\circ}$ and the redshifted lobe in the south at a PA of $\sim$195$^{\circ}$, and Per-emb-55 is associated with a northwest-southeast bipolar outflow with the blueshifted lobe in the south at a PA of $\sim$115$^{\circ}$ and the redshifted lobe in the north at a PA of $\sim$295$^{\circ}$ \citep{2016ApJ...820L...2L,2017ApJ...846...16S}. 
In addition to these outflows, the $^{12}$CO emission exhibits complex structures on a $\sim$8000 au scale around Per-emb-8 and 55. The C$^{18}$O emission also shows an asymmetric morphology centered at Per-emb-8 with an extension toward Per-emb-55 \citep{2019ApJS..245...21S}. 
On a smaller scale of $\sim$300 au, warped structures are observed around Per-emb-8 in the 1.3 mm continuum emission with ALMA at a resolution of $0\farcs27 \times 0\farcs17$ (81 au $\times$ 51 au), suggesting that Per-emb-8 is surrounded by a warped protostellar envelope or disk \citep{Tobin...et...al.2018ApJ...867...43T}.
These features may be related to the interaction between the companions or the turbulent environments, while the detailed gas kinematics and distributions remain unresolved. 
Therefore, this young triple system is an excellent target for further investigation on gas dynamics within proto-multiple systems.

To study how stars form in multiple systems, we have analyzed the archival ALMA 1.3 mm continuum, $^{12}$CO (2--1), and C$^{18}$O (2--1) data of Per-emb-8 and 55, which has a high resolution of 0\farcs4 to further resolve their gas kinematics and distributions. This paper is organized as follows. Section~\ref{sec:observation} describes the ALMA observations and data reduction. Section~\ref{sec:results} presents the results and is separated into subsections about the 1.3 mm continuum, $^{12}$CO line, and C$^{18}$O line emission. Section~\ref{sec: analysis} presents our analysis of the velocity structures, the protostellar masses in the system, and the warped accretion flows. Section~\ref{sec:discussion} discusses the possible origins of the observed features, and Section~\ref{sec: summary} summarizes our results.

\section{Observations} \label{sec:observation}
We retrieved the ALMA 1.3 mm continuum, $^{12}$CO (2-1), and C$^{18}$O (2-1) data of Per-emb-8 and Per-emb-55 from the archive (Project ID: 2017.1.01078.S, PI: Dominique Segura-Cox). The continuum data has been analyzed in \cite{2020A&A...640A..19T} to study the disk mass. The observations were conducted with 44 to 48 antennas from 2018 September 22 to 2018 September 24. The baseline lengths range from 14 m to 1397 m, and the maximum recoverable angular scale is $\sim$4$\farcs$ The pointing center is R.A (J2000) = 3$^h$44$^m$43$^s$.982, and decl. (J2000) = 32$^{\circ}$1$^{\prime}$35$\farcs$209. The calibration of the raw visibility data was done with the pipeline of the Common Astronomy Software Applications (CASA) version 5.4.0. Then we performed self calibration on the phase of the continuum data and applied the solution to the molecular-line data.
We subtracted the continuum from the molecular line data using the CASA task "$uvcontsub$". We generated continuum and line images using the CASA task "$tclean$" with the Briggs robust parameter of +0.5. The channel widths of the image cubes are 0.4 km s$^{-1}$ for the $^{12}$CO emission and 0.1 km s$^{-1}$ for the C$^{18}$O emission. The synthesized beams are 0$\farcs$39 $\times$ 0$\farcs$27 for the 1.3 mm continuum image, 0$\farcs$40 $\times$ 0$\farcs$27 for the $^{12}$CO image cube, and 0$\farcs$42 $\times$ 0$\farcs$29 for the C$^{18}$O image cube. The rms noise levels are $\sim$0.05 mJy beam$^{-1}$ in the 1.3 mm continuum image, $\sim$3 mJy beam$^{-1}$ per channel in the $^{12}$CO image cube, and $\sim$5 mJy beam$^{-1}$ per channel in the C$^{18}$O image cube.

\section{Results} \label{sec:results}

\subsection{1.3 mm Continuum Emission} \label{subsec:continuum}

\begin{figure*}
    \centering
    \includegraphics[width = 0.65\linewidth]{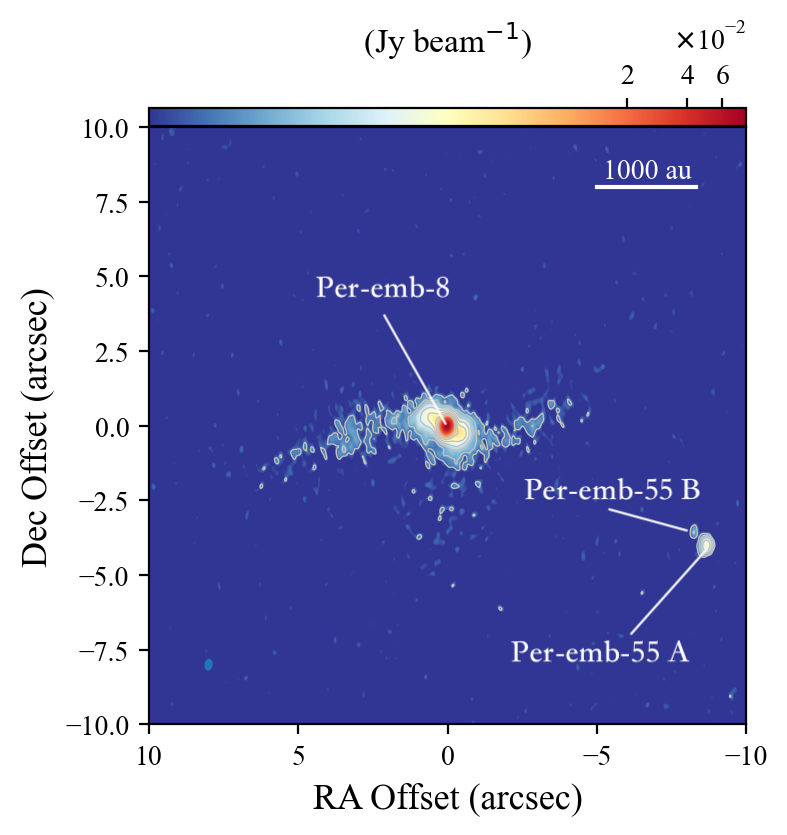}
    \caption{1.3 mm continuum image of Per-emb-8 and Per-emb-55 obtained with the ALMA observations. The source at the center is Per-emb-8. The two compact sources on the lower right side are Per-emb-55 $A$ and $B$. The projected separation between Per-emb-8 and 55 is 2782 au.  Contour levels are 3$\sigma$, 6$\sigma$, 12$\sigma$, 24$\sigma$, 48$\sigma$, 96$\sigma$, 192$\sigma$, where 1$\sigma$ is 0.05 mJy beam$^{-1}$. The filled ellipse in the lower left corner denotes the beam size.}
    \label{img_continuum}
\end{figure*}

Figure \ref{img_continuum} shows the 1.3 mm continuum image of Per-emb-8 and Per-emb-55 on a scale of 20$\arcsec$ (6000 au). 
In this image, the continuum source at the center is Per-emb-8, and Per-emb-55 is located at the lower right corner. 
In Per-emb-55, the southern component is source $A$, and the northern component is source $B$. 

The continuum emission in Per-emb-8 shows a central component elongated from northeast to southwest with an apparent size of $\sim$700 au, and two arm-like structures elongated along the east-west direction with a length of $\sim$1000 au, connecting the western and eastern parts of the central component.
To measure the size and flux of the continuum emission, we performed fitting of model intensity distributions to the continuum data. 
The 1.3 mm continuum emission has been resolved with ALMA at a resolution of 0$\farcs$27$\times$0$\farcs$17 \citep{Tobin...et...al.2018ApJ...867...43T}, and it shows a warped disk with an outer disk along a PA of 57.5$^{\circ}$$\pm1.6^{\circ}$ and an inner compact component with a deconvolved full width half maximum (FWHM) size of 0$\farcs$16 $\times$ 0$\farcs$094 along the east-west direction at a PA of 120.9$^{\circ}$$\pm2.6^{\circ}$. 
With more short-baseline data, we expect to better recover the emission from the outer disk.

\begin{figure*}
    \centering
    \includegraphics[width = \linewidth]{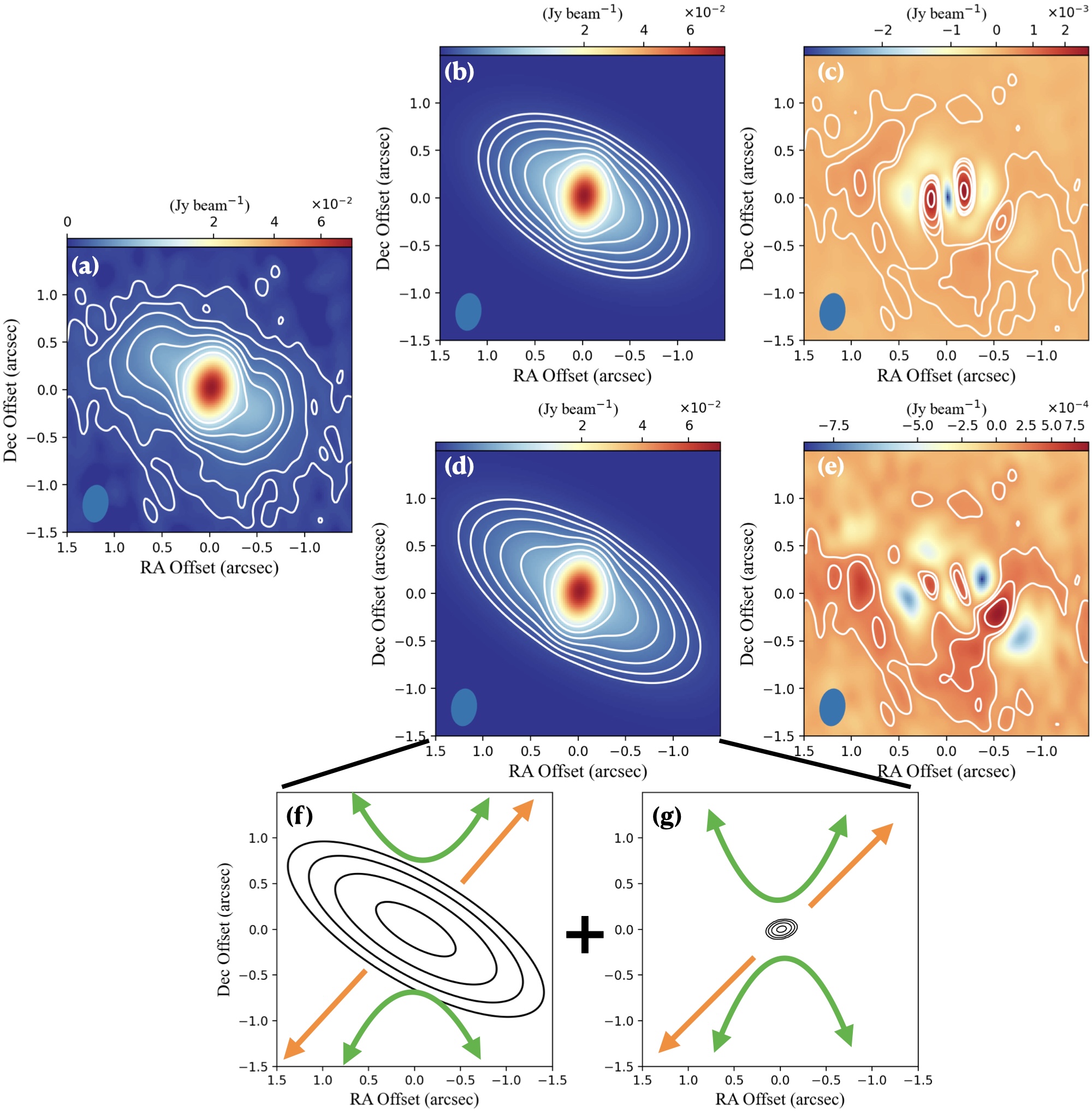}
    \caption{To understand the intensity distribution of the central continuum emission of Per-emb-8, we fitted it with the Gaussian function. (a) Zoom-in of the 1.3 mm continuum image of Per-emb-8. (b) The model image of the 1.3 mm continuum emission of Per-emb-8. The model consists of a point source and a Gaussian component. (c) The residual map after subtracting the model image of a point source and a Gaussian component from the observed image. (d) The model image consists of two Gaussian components. The deconvolved PA of the major axes of the extended and compact Gaussian components are 59$^{\circ}$ and 103$^{\circ}$, respectively. (e) The residual map after subtracting the model image of two Gaussian components from the observed image. A filled ellipse in the bottom left corner of each panel denotes the beam size. Contour levels start from 3$\sigma$ to 192$\sigma$ in steps of increasing by a factor of two, where 1$\sigma$ is 0.05 mJy beam$^{-1}$. (f) and (g) show the two Gaussian components of the model presented in (d). Contour levels are 1$\%$, 4$\%$, 16$\%$, and 64$\%$ of peak intensity. The green and orange arrows represent the two outflow directions (see Sec.~\ref{subsec:12CO}).}
    \label{img_Per8_cont_fitting}
\end{figure*}

We first fitted the continuum emission with a Gaussian function and a point source, since the inner disk is smaller than half of our angular resolution, which may not be resolved in our data.
The fitting result is shown in Fig.~\ref{img_Per8_cont_fitting}b. The continuum peak position of Per-emb-8 is measured to be R.A. (J2000) = 3$^h$44$^m$43$^s$.982, decl. (J2000) = 32$^{\circ}$1$\arcmin$35$\farcs$20.
There are strong positive residuals at a level of 15$\sigma$ near the center at offsets of 0$\farcs$2 along the northwest-southeast direction. 
To obtain a better fit, we adopted two Gaussian components in our model (Fig.~\ref{img_Per8_cont_fitting}d). This fitting result suggests that the continuum emission is better represented by two Gaussian functions (Fig.~\ref{img_Per8_cont_fitting}e), an extended component with a deconvolved FWHM size of 1$\farcs$22 $\times$ 0$\farcs$45 along the northeast-southwest direction at a PA of 59$^{\circ}$$\pm0.1^{\circ}$ and an inner compact component with a deconvolved FWHM size of 0$\farcs$13 $\times$ 0$\farcs$08 along the east-west direction at a PA of 103$^{\circ}$$\pm0.2^{\circ}$. 
The orientations of our fitted two Gaussian components are consistent with those of the inner and outer disks within $\sim$10--20$^{\circ}$ reported in \cite{Tobin...et...al.2018ApJ...867...43T}. The fitted total flux of the two Gaussian components is 236 mJy, which is $\sim$2 times higher than that measured by \cite{Tobin...et...al.2018ApJ...867...43T}. 

\begin{figure*}
    \centering
    \includegraphics[width = \linewidth]{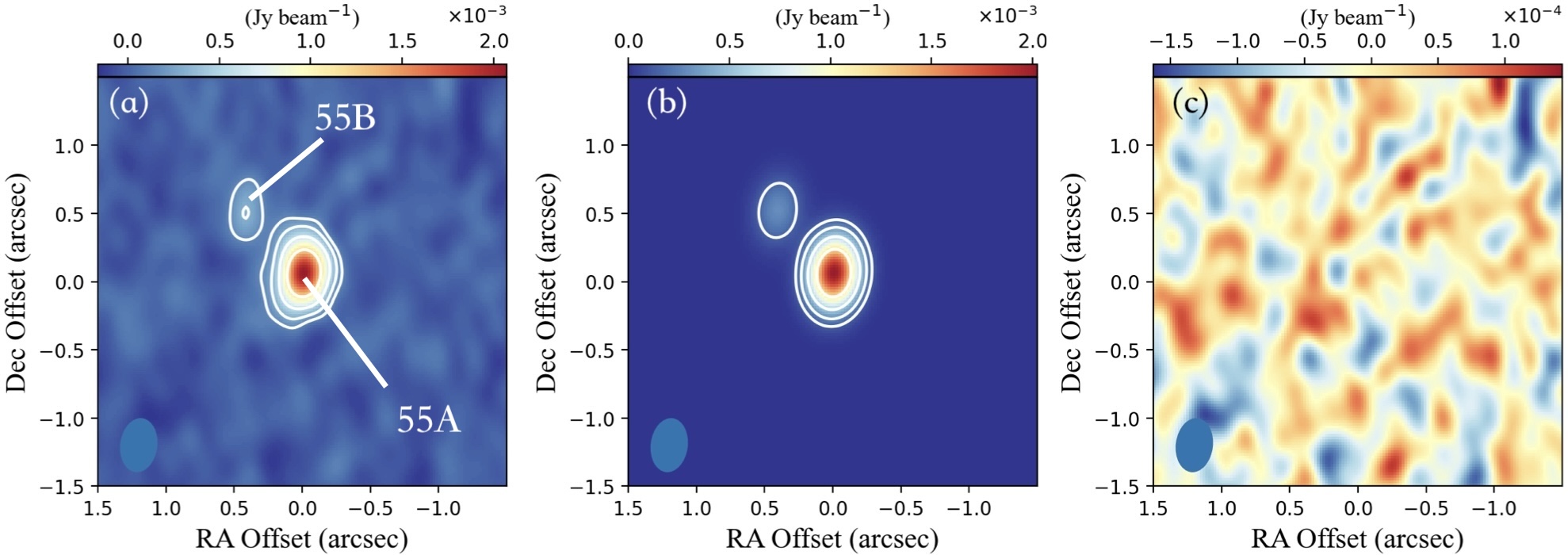}
    \caption{(a) Zoom-in of the 1.3 mm continuum image of Per-emb-55. (b) The model image of the 1.3 mm continuum emission of Per-emb-55. The model consists of a Gaussian intensity distribution with the deconvolved PA of major axis of $\sim154^{\circ}$ for source $A$ and a point source for source $B$. (c) The residual map that after subtracting the model image from the observed image. A filled ellipse in the bottom left corner of each panel denotes the beam size. The contour levels start from 3$\sigma$ to 192$\sigma$ in steps of increasing by a factor of two, where 1$\sigma$ is 0.05 mJy beam$^{-1}$.}
    \label{img_Per55_cont_fitting}
\end{figure*}

The continuum emission of Per-emb-55 $A$ shows a compact blob elongated along the north-south direction (Fig.~\ref{img_Per55_cont_fitting}a). 
We fitted a 2D Gaussian to it to measure its size and flux.
The fitted Gaussian component has a total flux of 3.6 mJy and a deconvolved FWHM size of 0$\farcs09$$\times$0$\farcs03$ with a PA of 154$^{\circ}$$\pm 3.4 ^{\circ}$. The continuum peak position of source $A$ is at R.A. (J2000) = 3$^h$44$^m$43$^s$.301, decl. (J2000) = 32$^{\circ}$1$\arcmin$31$\farcs$21.
The apparent size of the continuum emission of source $B$ is comparable to the beam size of 0$\farcs$4$\times$0$\farcs$3, so it is not resolved with the observations. We fitted the continuum emission of source $B$ with a point source. The total flux of Per-emb-55 $B$ is measured to be 0.5 mJy, and the continuum peak position is measured to be R.A. (J2000) = 3$^h$44$^m$43$^s$.334, decl. (J2000) = 32$^{\circ}$1$\arcmin$31$\farcs$65. 

Under the assumption that the dust continuum emission is optically thin, the mass of the circumstellar material can be estimated as
\begin{equation} \label{Mass_dust}
    M_d = \frac{F_{1.3 mm}d^2}{\kappa_{1.3 mm}B(T_{dust})},
\end{equation}
where $F_{1.3mm}$ is the total flux at 1.3 mm, $d$ is the distance, $\kappa_{1.3mm}$ is the dust mass opacity at 1.3 mm, and $T_{dust}$ is the dust temperature. 
We adopt $\kappa_{1.3mm}$ = 0.1 $\times$ (0.3 mm/$\lambda$)$^{\beta}$ cm$^2$ g$^{-1}$ = 0.023 cm$^2$ g$^{-1}$ with $\beta$ = 1 \citep{1990AJ.....99..924B}. The dust temperature is estimated with
\begin{equation} \label{T_dust}
    T_{dust}=T_0\left(\frac{L_{bol}}{L_{\odot}}\right)^{0.25},
\end{equation}
where $T_0$ is 43 K, and $L_{bol}$ = 2.6 $\pm$ 0.5 $L_{\odot}$ for Per-emb-8 and $L_{bol}$ = 1.8 $\pm$ 0.8 $L_{\odot}$ for Per-emb-55 \citep{2016ApJ...818...73T, 2020ApJ...890..130T}.
The mass of the circumstellar material around Per-emb-8 is estimated to be 0.055$\pm$0.003 $M_{\odot}$ with a dust temperature of 54$\pm$3 K, 
and those around Per-emb-55 $A$ and $B$ are estimated to be $(9\pm1) \times 10^{-4}$ $M_{\odot}$ and $(1.3\pm0.2) \times 10^{-4}$ $M_{\odot}$ with the $T_{dust} = 48.5\pm5.5$ K, respectively. If $T_{dust}$ is adopted to be 30 K, the masses of the circumstellar material around Per-emb-8 and Per-emb-55 $A$ and $B$ are estimated to be 0.11 $M_{\odot}$, $1.6\times 10^{-3}$ $M_{\odot}$, $2.3\times 10^{-4}$ $M_{\odot}$, respectively.

% \begin{figure*}
%     \centering
%     \includegraphics[width = \linewidth]{Fig_Per-emb-55 continuum fitting.jpeg}
%     \caption{(a) Zoom-in of the 1.3 mm continuum image of Per-emb-55. (b) The model image of the 1.3 mm continuum emission of Per-emb-55. The model consists of a Gaussian intensity distribution with the deconvolved PA of major axis of $\sim154^{\circ}$ for source $A$ and a point source for source $B$. (c) The residual map that after subtracting the model image from the observed image. A filled ellipse in the bottom left corner of each panel denotes the beam size. The contour levels start from 3$\sigma$ to 192$\sigma$ in steps of increasing by a factor of two, where 1$\sigma$ is 0.05 mJy beam$^{-1}$.}
%     \label{img_Per55_cont_fitting}
% \end{figure*}

\subsection{$^{12}$CO (2-1) Emission\label{subsec:12CO}}

\begin{figure*}
    \centering
    \includegraphics[width = \linewidth]{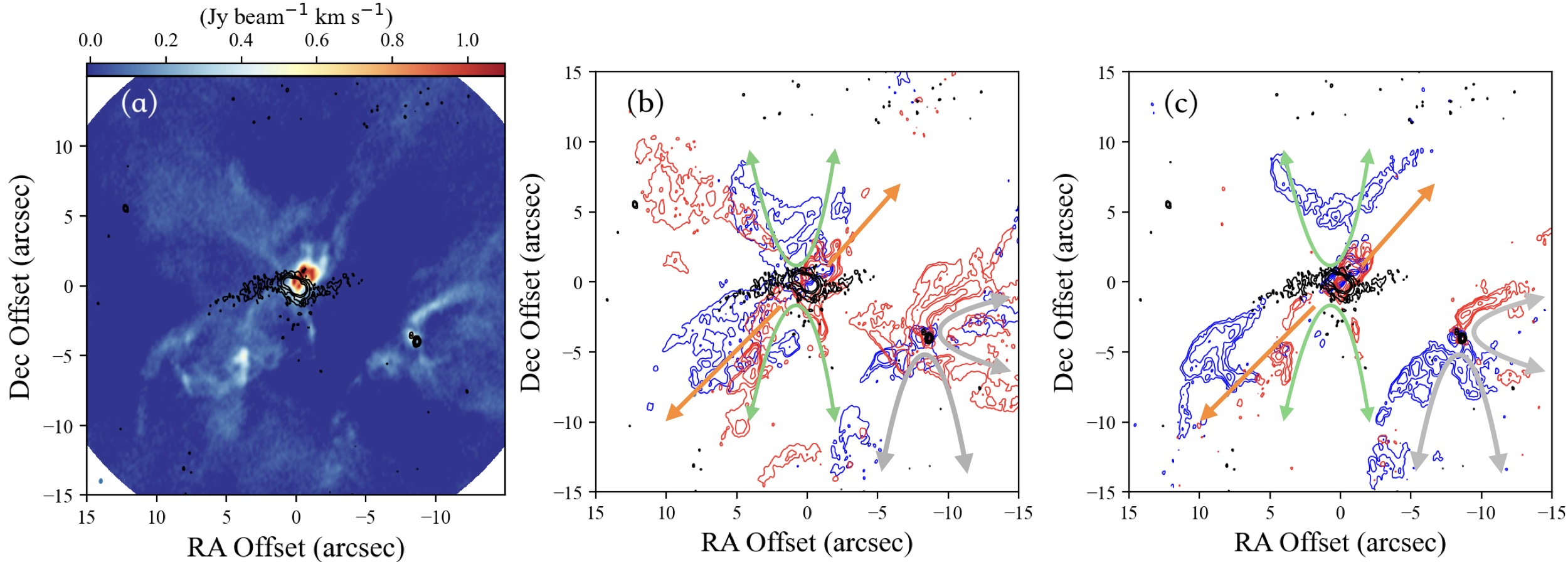}
    \caption{(a) 1.3 mm continuum emission (contour) overlaid on the moment 0 map (color) of the $^{12}$CO emission. A filled ellipse in the bottom left corner denotes the beam size. (b) Moment 0 map of the low velocity $^{12}$CO emission. Blue and red contours show the integrated blueshifted and redshifted emission, respectively. The integrated velocity ranges are $V_{LSR}$ = 7.2 to 8.4 km s$^{-1}$ for the blueshifted emission and $V_{LSR}$ = 11.4 to 13.2 km s$^{-1}$ for the redshifted emission, where the systemic velocity is 10.3 km s$^{-1}$. The green arrows delineate the cavity walls of the north-south bipolar outflow, and the orange arrows denote the directions of the northwest-southeast bipolar outflow, associated with Per-emb-8. The gray arrows delineate the cavity walls of the misaligned outflows toward the west and south associated with Per-emb-55. (c) Moment 0 map of the high velocity $^{12}$CO emission. The integrated velocity ranges are $V_{LSR}$ = 4.4 to 6.8 km s$^{-1}$ and $V_{LSR}$ = 13.6 to 17.6 km s$^{-1}$ for the blueshifted and redshifted emission, respectively. A filled ellipse in the bottom left corner of each panel denotes the beam size. The contour levels are from 3$\sigma$ to 48$\sigma$ in steps of increasing by a factor of two, where 1$\sigma$ is 6.9 mJy beam$^{-1}$ km s$^{-1}$.}
    \label{img_12CO_mom0}
\end{figure*}

Figure~\ref{img_12CO_mom0} shows the moments 0 map of the $^{12}$CO (2-1) emission. The red and blue contours present the redshifted and blueshifted emission, respectively. Fig.~\ref{img_12CO_mom0}(b) shows the $^{12}$CO (2-1) emission at the low velocities ($V_{LSR}$ = 7.2 to 8.4 km s$^{-1}$ and $V_{LSR}$ = 11.4 to 13.2 km s$^{-1}$. Fig.~\ref{img_12CO_mom0}(c) shows the $^{12}$CO (2-1) emissions at the high velocities ($V_{LSR}$ = 4.4 to 6.8 km s$^{-1}$ and $V_{LSR}$ = 13.6 to 17.6 km s$^{-1}$). The systemic velocity of the Per-emb-8 and 55 system is estimated to be 10.3 km s$^{-1}$ (Sec.~\ref{subsec: Kepl. Disk and Mass}). In these figures, the $^{12}$CO emission exhibits a complex morphology around Per-emb-8. The bipolar outflow associated with Per-emb-8 has been reported by the MASSES survey with the blue- and redshifted lobes at PA of $\sim$15$^{\circ}$ and $\sim$195$^{\circ}$, respectively \citep{2017ApJ...846...16S}.

\begin{figure*}
    \centering
    \includegraphics[width = 0.86\linewidth]{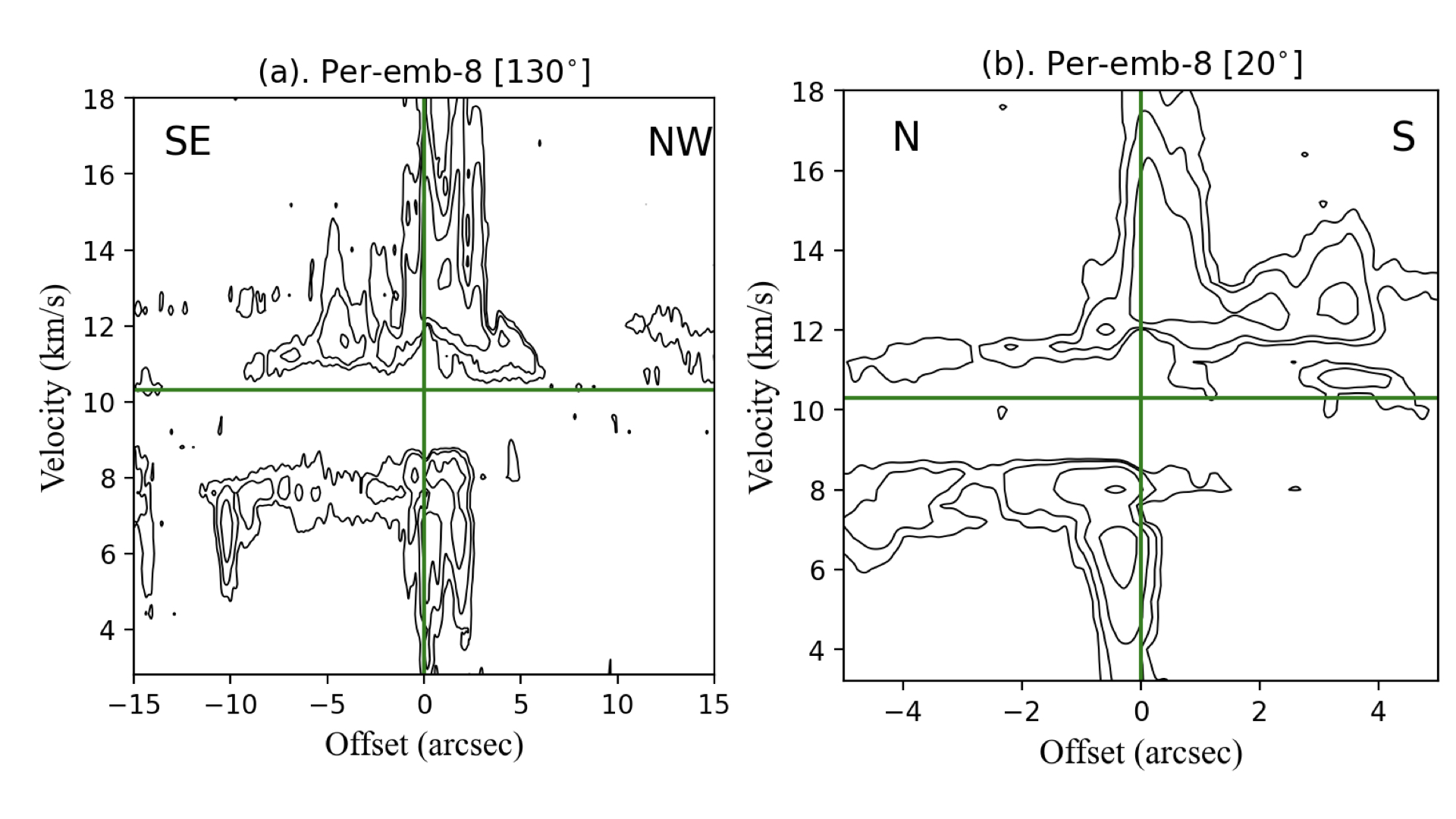}
    \caption{P-V diagrams of the $^{12}$CO emission along the outflow directions at a PA of 130$^{\circ}$ (a), and a PA of 20$^{\circ}$ (b) in Per-emb-8. The green vertical and horizontal lines denote the protostellar position and the systemic velocity of 10.3 km s$^{-1}$, respectively. Contour levels are from 3$\sigma$ to 81$\sigma$ in steps of increasing by a factor of three.}
    \label{img_PV_Per8_outflow}
\end{figure*}

In Fig.~\ref{img_12CO_mom0}, at both low and high velocities, blue- and redshifted V-shaped structures extending to the north and south around Per-emb-8 are observed, respectively. At the low velocities, the apexes of these V-shaped structures are approximately coincident with the position of Per-emb-8. At the high velocities, the northern blueshifted V-shaped structure is shifted further toward the north. The central axes of these V-shaped structures are not perpendicular to the elongation of the extended components of the central continuum emission in Per-emb-8.
Fig.~\ref{img_PV_Per8_outflow} (b) shows the P-V diagram of the $^{12}$CO emission centered at Per-emb-8 along a PA of $20^{\circ}$, which passes through the northern and southern V-shaped structures and is also the outflow axis reported in the SMA study \citep{2017ApJ...846...16S}. The central $^{12}$CO component at offsets smaller than 1$\arcsec$ exhibits large line widths of $>$5--6 km s$^{-1}$ at both blue- and redshifted velocities, and is likely associated with the disk (see Sec.~\ref{subsec: Kepl. Disk and Mass}). At offsets larger than 2$\arcsec$, the velocities of the $^{12}$CO emission increase as the offsets increase. This velocity profile is expected in outflow material swept up by wind or jets \citep{2007prpl.conf..245A}, and such a feature has been observed in other protostellar outflows \citep[e.g.,][]{2019ApJ...883....1Z}.
Thus, these V-shaped structures toward the north and south most likely trace the walls of the outflow cavities, and this outflow direction is consistent with that of the outflow identified with the SMA observations \citep{2017ApJ...846...16S}.

Additionally, there is $^{12}$CO emission extending to the southeast at blueshifted velocities and to the northwest at both blue- and redshifted velocities around Per-emb-8. These extensions are more clearly seen at the high velocities, and the northwestern extension is brighter and exhibits strong intensity peaks at 2$\arcsec$--4$\arcsec$ from the central protostar. Their orientation is approximately perpendicular to the elongation of the extended component of the central continuum emission in Per-emb-8. 
Fig.~\ref{img_PV_Per8_outflow} (a) shows the P-V diagram of the $^{12}$CO emission along a PA of 130$^{\circ}$, perpendicular to the elongation of the central continuum emission and passing through the northwest--southeast $^{12}$CO extensions. 
The northwestern extension has a large line width of $>$4--6 km s$^{-1}$ at both redshifted and blueshifted velocities. 
The southeastern extension exhibits several compact $^{12}$CO components with large line widths of 3--4 km s$^{-1}$ in the P-V diagram, and the components further away from the protostar have larger maximum velocities with respect to the systemic velocity. These velocity features are different from those in the north-south outflows associated with Per-emb-8, and they are similar to the theoretical velocity profiles of outflows driven by jet bow shocks \citep{2007prpl.conf..245A}. These results suggest that there is another bipolar outflow along the northwest-southeast direction at a PA of $\sim$130$^{\circ}$ associated with Per-emb-8. 
This northwest-southeast $^{12}$CO bipolar outflow was not clearly identified in the previous SMA observations, possibly due to their lower resolution of 2$\farcs$8$\times$2$\farcs$5.

\begin{figure*}
    \centering
    \includegraphics[width = 0.86\linewidth]{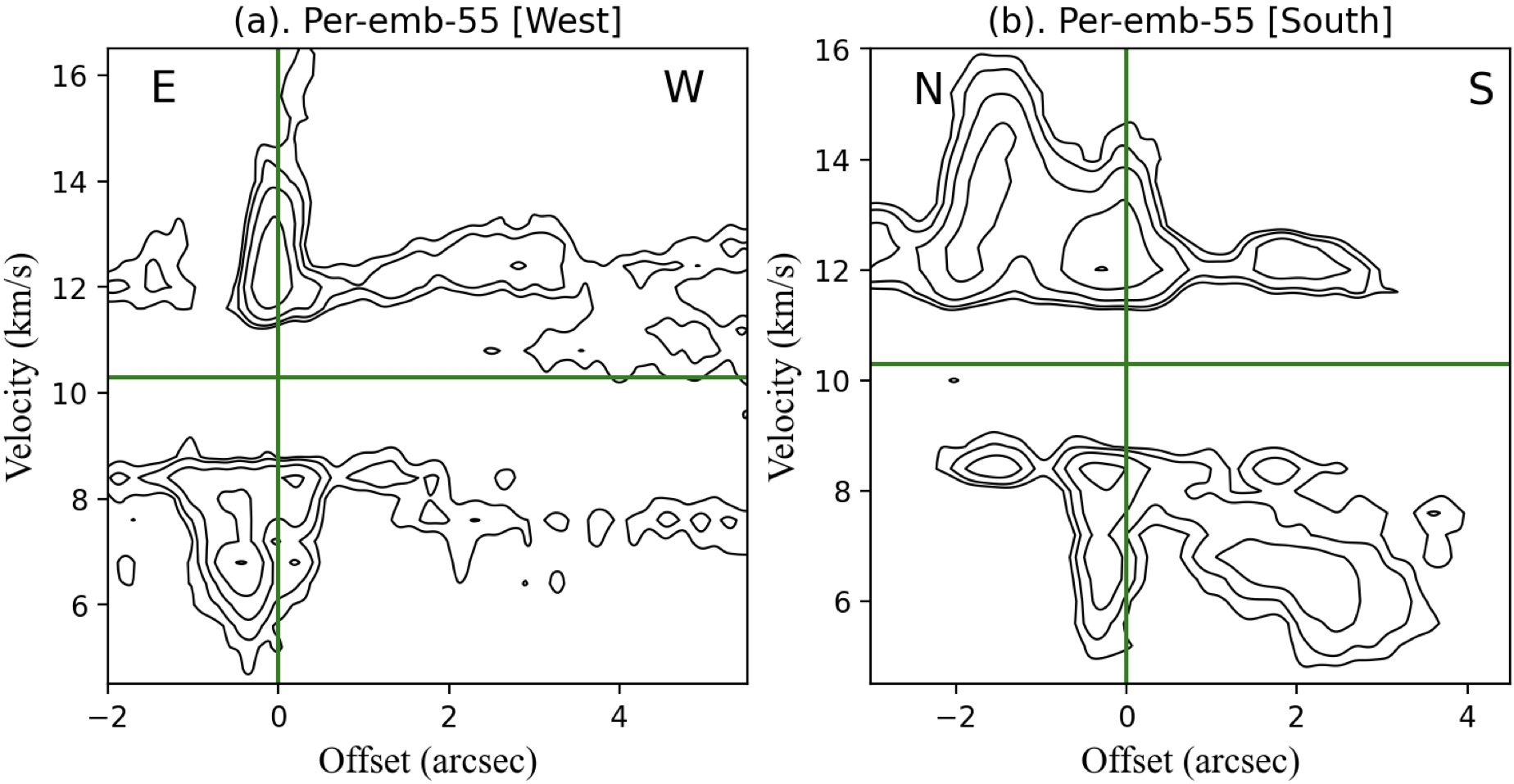}
    \caption{P-V diagrams of the $^{12}$CO emission along the central axes of the redshifted and blueshifted V-shaped outflows at PA of -90$^{\circ}$ (a) and 180$^{\circ}$ (b) in Per-emb-55, respectively. The green vertical and horizontal lines denote the position of Per-emb-55 $A$ and the systemic velocity of 10.3 km s$^{-1}$, respectively.  Contour levels are from 3$\sigma$ to 48$\sigma$ in steps of increasing by a factor of two. }
    \label{img_PV_Per55_outflow}
\end{figure*}

Around Per-emb-55, the presence of a bipolar outflow has been reported by the MASSES survey with its blue- and redshifted lobe at a PA of $\sim$115$^{\circ}$ and $\sim$-65$^{\circ}$, respectively. In our $^{12}$CO data, clear blue- and redshifted V-shaped structures to the south and west are observed, respectively (Fig.~\ref{img_12CO_mom0}). The directions of the blue- and redshifted V-shaped structures are misaligned. 
These directions are inconsistent with the outflow directions of 115$^{\circ}$ for the blueshifted lobe and -65$^{\circ}$ for the redshifted lobe reported in \cite{2017ApJ...846...16S}. This is because the V-shaped morphologies are better resolved with the ALMA data having the higher resolution. 
Figure \ref{img_PV_Per55_outflow} presents the P-V diagrams of the $^{12}$CO emission along the central axes of the V-shaped structures at PA of -90$^{\circ}$ and 180$^{\circ}$. 
The $^{12}$CO emission around zero offset with a large line width of 2--4 km s$^{-1}$ is likely associated with the protostellar disks or inner protostellar envelopes of Per-emb-55. 
In both P-V diagrams (Fig.~\ref{img_PV_Per55_outflow}), the $^{12}$CO emission at positive offsets of 1$\arcsec$--4$\arcsec$ exhibits higher velocities at outer radii, meaning that the velocities increase toward the east and south. 
Thus, the morphologies and velocity structures of these V-shaped $^{12}$CO structures toward the east and south are consistent with the expectation for swept-up outflow material, 
and they most likely trace the walls of the outflow cavities associated with Per-emb-55. 
The previous SMA observations mainly detected the northern part of the redshifted outflow cavity wall and the eastern part of the blueshifted outflow cavity wall. 

\subsection{C$^{18}O$(2-1) Emission} \label{subsec:C18O}

\begin{figure*}
    \centering
    \includegraphics[width = \linewidth]{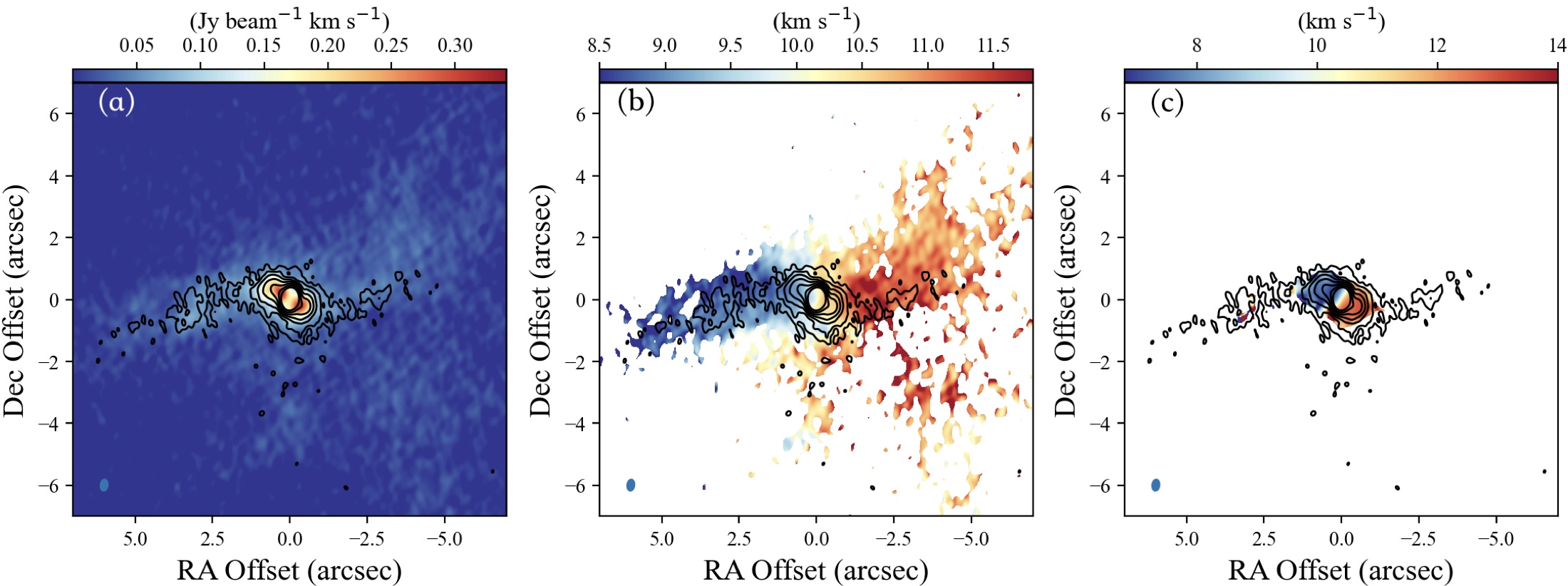}
    \caption{(a) Moment 0 of the C$^{18}$O emission in Per-emb-8. (b) Moment 1 map of the C$^{18}$O emission. The integrated velocity range is from 8.3 km s$^{-1}$ to 11.7 km s$^{-1}$ to emphasize the velocity features of the warped structures. (c) Moment 1 map of the C$^{18}$O emission in the central region around Per-emb-8. The integrated velocity range is from 5.1 km s$^{-1}$ to 15.0 km s$^{-1}$. A filled ellipse in the bottom left corner of each panel denotes the beam size. Contours show the 1.3 mm continuum emission, the same in Fig.~\ref{img_continuum}}
    \label{img_C18O_mom0_1}
\end{figure*}

Figure \ref{img_C18O_mom0_1} presents the moment 0 and 1 maps of the C$^{18}$O emission overlaid with the 1.3 mm continuum map of Per-emb-8. We do not detect C$^{18}$O emission toward Per-emb-55.
The moment 0 map (Fig.~\ref{img_C18O_mom0_1}a) shows that the central compact emission exhibits two intensity peaks around the center, and it is elongated along the major axis of the continuum emission.
Besides, the C$^{18}$O emission is extended towards the east and west from the central component, 
and there is also fainter C$^{18}$O emission in the south. 
The eastern and western extensions of the C$^{18}$O emission form a warped morphology and are coincident with the arm-like structures observed in the continuum emission.

The moment 1 map (Fig.~\ref{img_C18O_mom0_1}c) shows that the central component of the C$^{18}$O emission exhibits a clear velocity gradient along the major axis of the continuum emission, where the northeastern side is blueshifted and the southwestern side is redshifted. 
This dominant velocity gradient along the elongation of the central continuum component around the protostar hints at disk rotation.

To emphasize the velocity features of the C$^{18}$O emission extending toward the east and west, we integrated the C$^{18}$O emission at velocities from $V_{LSR}$ = 8.3 km s$^{-1}$ to 11.7 km s$^{-1}$ and generated a moment 1 map, shown in Fig.~\ref{img_C18O_mom0_1}(b). 
The extended warped C$^{18}$O emission exhibits a clear velocity gradient along the east-west direction with the redshifted emission in the west and blueshifted emission in the east.

\section{Analysis} \label{sec: analysis}

\subsection{Keplerian Disks and Protostellar Masses} \label{subsec: Kepl. Disk and Mass}

\begin{figure*}
    \centering
    \includegraphics[width = 0.86\linewidth]{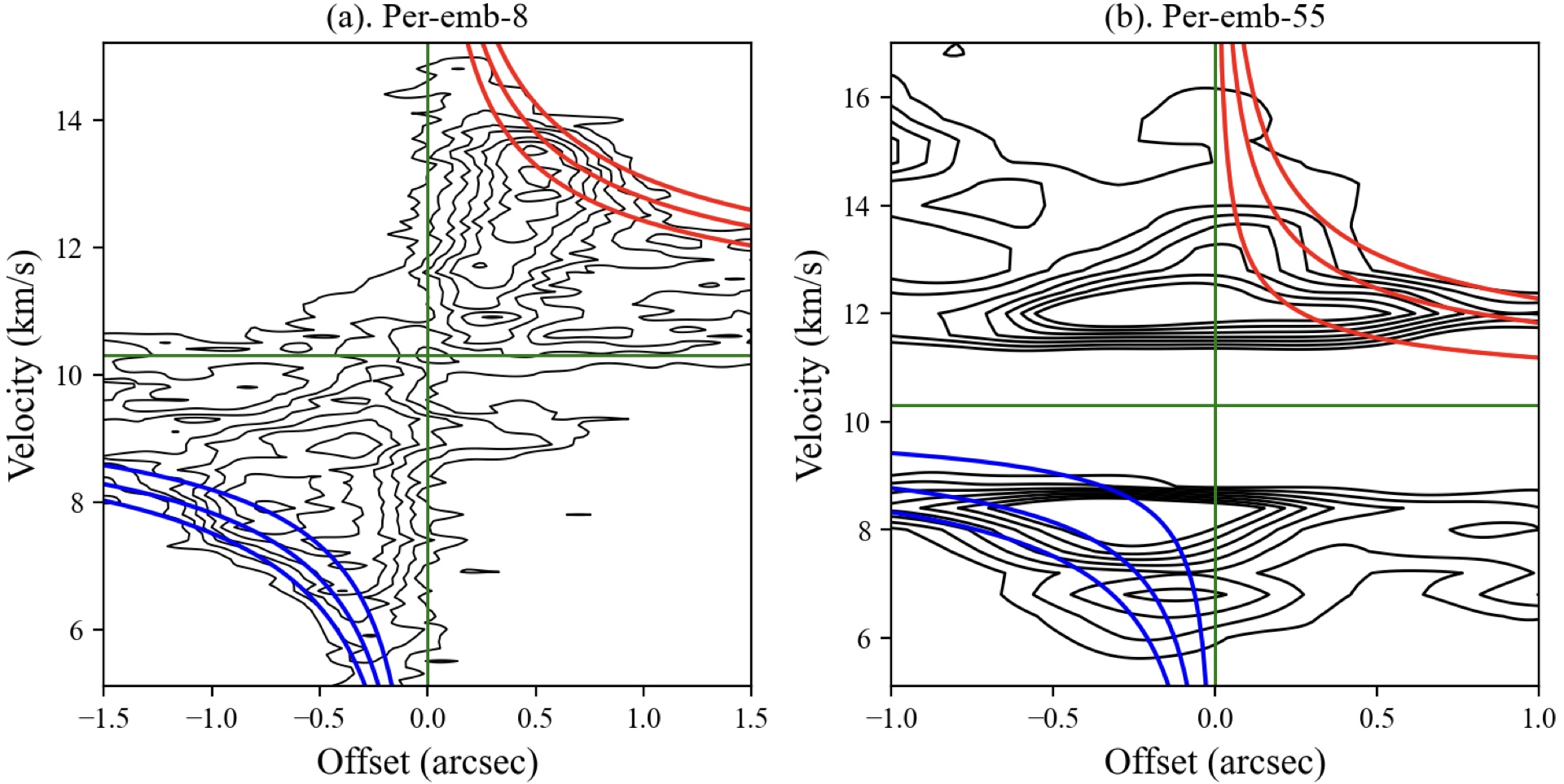}
    \caption{(a) P-V diagram of the C$^{18}$O emission along the major axis of the continuum emission in Per-emb-8. The blue and red curves show the velocity profiles of the Keplerian rotation around 1.6, 2.2, and 2.8 M$_\odot$ under the assumption that the inclination angle of the disk plane is 65$^{\circ}$. (b) P-V diagram of the $^{12}$CO emission along the major axis of the continuum emission in Per-emb-55 $A$. The blue and red curves show the velocity profiles of the Keplerian rotation around 0.2, 0.6, and 1 M$_\odot$ under the assumption that the inclination angle of the disk plane is 75$^{\circ}$. Contour levels are from 3$\sigma$ to 21$\sigma$ in steps of 3$\sigma$, where 1$\sigma$ is 5 mJy beam$^{-1}$ for the C$^{18}$O emission and 3 mJy beam$^{-1}$ for the $^{12}$CO emission.}
    \label{img_PV_disk}
\end{figure*}

The extended component of the central continuum emission around Per-emb-8 is elongated perpendicular to the northwest-southeast bipolar outflow, and it exhibits the signature of a dominant rotational motion observed in the C$^{18}$O emission. 
Thus, this central component likely traces the protostellar disk around Per-emb-8.
To further characterize its velocity feature, 
we applied the Python package $SLAM$ \citep{2023zndo...7783868A}, which measures the rotational velocity as a function of radius and fits the profile with a power-law function.
Figure \ref{img_PV_disk} (a) presents the P-V diagram of the C$^{18}$O emission along the major axis of the continuum emission. 
In the P-V diagram, the velocities increase as the radii decrease. This profile can be explained with Keplerian rotation. 
By fitting the Keplerian velocity profile to the edge and ridge of the observed P-V diagram using $SLAM$ with the MCMC method and setting the emission threshold at 3$\sigma$ noise level, the stellar mass of Per-emb-8 is estimated to be 1.6--2.8 $M_{\odot}$ at an inclination angle of 65$^{\circ}$, and the systemic velocity is estimated to be 10.3 km s$^{-1}$. The inclination angle is measured from the length ratio of the deconvolved major and minor axes of the disk in the continuum emission. 
We note that the central continuum emission around Per-emb-8 consists of two components and possibly traces the warped disk. The size of the inner component is comparable to the angular resolution of the C$^{18}$O data, so its velocity structure cannot be resolved and does not affect our fitting results.

Since the C$^{18}$O emission is undetected toward Per-emb-55, we investigate the $^{12}$CO emission to estimate its stellar mass. Figure~\ref{img_PV_disk} (b) shows the P-V diagram of the $^{12}$CO emission along the major axis of the continuum emission in Per-emb-55 $A$. There is no clear disk-like component associated with Per-emb-55 $A$ seen in the $^{12}$CO moment 0 map (Fig.~\ref{img_12CO_mom0}), possibly due to the contamination from the outflows. Nevertheless, the P-V diagram along the major axis of the continuum emission shows increasing velocities with decreasing radii within a radius of 0$\farcs$7 around Per-emb-55 $A$. This velocity feature is different from that of the outflows and can be the signature of disk rotation. With the limited resolution, this velocity profile is not well resolved due to the small disk around Per-emb-55 $A$ with an apparent diameter of $\lesssim$1$\arcsec$ in the continuum (Fig.~\ref{img_Per55_cont_fitting}), so $SLAM$ is not appliable.
We visually compared the observed P-V diagram with a series of Keplerian velocity profiles, shown as red and blue curves. Assuming that this velocity gradient traces the Keplerian rotation and the systemic velocity is the same as Per-emb-8, the stellar mass of Per-emb-55 $A$ is estimated to be 0.4--1 $M_{\odot}$ with the inclination angle of the disk plane of $75 ^{\circ}$ estimated from the aspect ratio of the continuum emission.
 
\begin{figure}
    \centering
    \includegraphics[width = 0.98\linewidth]{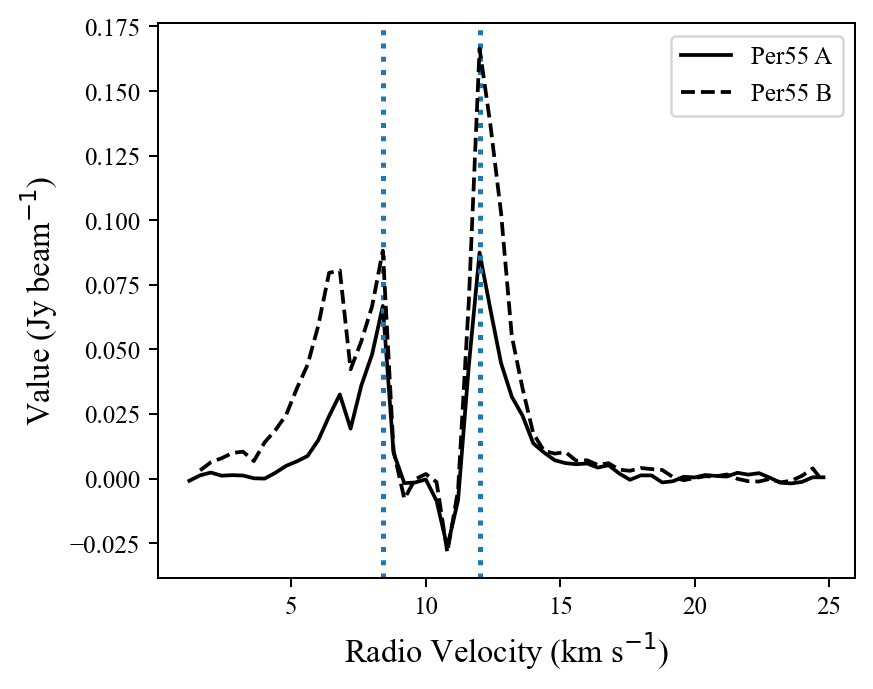}
    \caption{$^{12}$CO spectra of Per-emb-55 $A$ (solid line) and $B$ (dashed line). The blue dotted lines denote the velocities of 8.4 and 12 km s$^{-1}$. The spectra were extracted at the protostellar positions with aperture sizes comparable to the sizes of their 1.3 mm continuum emission above 3$\sigma$, which are $0.5\arcsec$ in radius for Per-emb-55 $A$ and $0.2\arcsec$ in radius for Per-emb-55 $B$, respectively.}
    \label{img_spectral_profile}
\end{figure}

 The disk of Per-emb-55 $B$ is not resolved in the continuum emission, and there is no feature of possible disk rotation observed in the lines, so we performed an order of magnitude estimate of its stellar mass based on the $^{12}$CO spectrum (Fig.~\ref{img_spectral_profile}).
 The $^{12}$CO spectrum at the protostellar position of Per-emb-55 $B$ shows a double peak line profile, similar to that of Per-emb-55 $A$, and the $^{12}$CO spectra of Per-emb-55 $A$ and $B$ have identical peak velocities. 
 Such double peak line profiles are often seen in protoplanetary disks due to disk rotation \citep{1998A&A...339..467G}. 
 Assuming that the double peak line profile is caused by Keplerian rotation but not missing flux, the radius of the disk is the same as the beam size ($\sim$60 au), 
the systemic velocity is the same as Per-emb-8, and the disk is edge on, the stellar mass of 
 Per-emb-55 $B$ is estimated to be 0.3 $M_{\odot}$. We note that this is only an order of magnitude estimate because there could be self absorption in the $^{12}$CO spectrum and the disk is not resolved. Therefore, the stellar mass of Per-emb-55 $B$ can be comparable to Per-emb-55 $A$, while Per-emb-8 is the primary star with its stellar mass dominant in this multiple protostellar system.

\subsection{Infalling Motion of the Warped Structures} \label{subsec: infalling motion}
We have identified arm-like warped structures connecting the Per-emb-8 disk extending toward the east and west in both the C$^{18}$O and continuum emission (Fig.~\ref{img_C18O_mom0_1}). 
In the moment 1 map (Fig.~\ref{img_C18O_mom0_1}b), the arm-like structures exhibit increasing velocities toward the central star, so they possibly trace infalling material toward the Keplerian disk.

\begin{figure*}
    \centering
    \includegraphics[width = \linewidth]{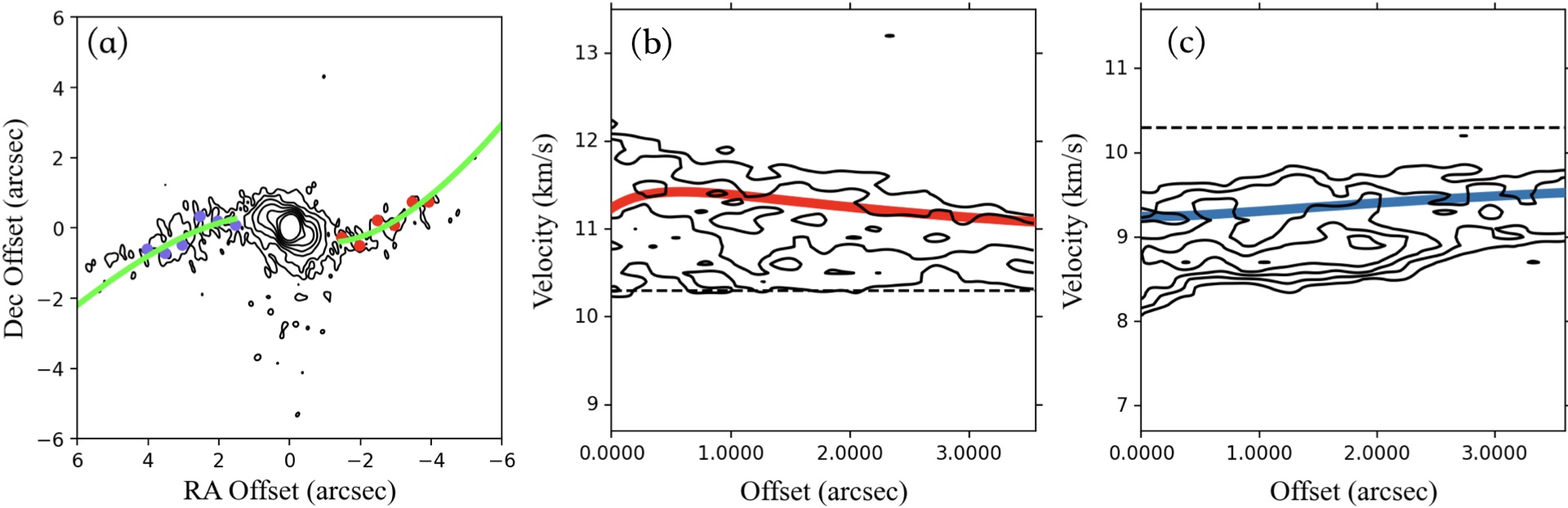}
    \caption{Comparison between the CMU model of infalling motion with the observed morphologies and velocity structures of the arm-like warped structures. (a) 1.3 mm continuum map of Per-emb-8 (contours) overlaid with the best-fit trajectories of the red- and blueshifted infalling flows from the CMU model (green curves). The continuum map is the same as in Fig.~\ref{img_continuum}. The red and blue circles represent the peak positions adopted for fitting of the trajectories of infalling motion. (b and c) P-V diagrams of the C$^{18}$O emission (contours) along the best-fit trajectories of the red- and blueshifted infalling flow. The red and blue lines show the line of slight velocities of infalling material along the best-fit trajectories from the CMU model. The dashed lines denote the systemic velocity of 10.3 km s$^{-1}$. Contours are from 3$\sigma$ to 15$\sigma$ in steps of 3$\sigma$.}
    \label{img_CMU_fitting}
\end{figure*}

To further understand the kinematics of the structures, we examined the spectra along the ridges of the warped structures. 
We extracted intensity profiles along the declination direction at right ascension offsets in steps of one beam size and measured the peak positions (blue and red circles in Fig.~\ref{img_CMU_fitting}). Then we calculated the trajectories of free-fall motion to fit these peak positions. On the assumption that the gravity is dominated by the central protostar, with a conserved angular momentum and zero total energy, the materials would be free-falling onto the disk and follow the parabolic trajectories \citep{1976ApJ...210..377U,1981Icar...48..353C}. These trajectories could result in the formation of the warped structures observed in the C$^{18}$O and continuum emission \citep[e.g.,][]{2014ApJ...793....1Y,2022ApJ...925...32T}. The parabolic trajectories of the infalling materials can be described as
\begin{equation}
    r = \frac{R_c\cos{\theta_0}\sin^2{\theta_0}}{\cos{\theta_0} - \cos{\theta}},
\end{equation}
where
\begin{equation}
    \cos{\theta} = \cos{\theta_0}\cos{\alpha},
\end{equation}
\begin{equation}
    \tan{(\phi - \phi_0)} = \tan{\alpha}/\sin{\theta_0}.
\end{equation}
In the above equations, $R_c$ is the centrifugal radius, and $\theta_0$ and $\phi_0$ are the initial polar and azimuthal angles of the infalling materials, and $\alpha$ is the angle between the initial and current position of the infalling materials measured from the protostellar position. 
In our calculation, the centrifugal radius is adopted to be the disk radius of 300 au, and $\theta_0$ and $\phi_0$ are adopted as free parameters. The results are shown in Fig.~\ref{img_CMU_fitting} (left). The morphologies of the redshifted and blueshifted warped structures can be explained with the parabolic trajectories of infalling materials with $\theta_0$ = 49$^{\circ}$ and $\phi_0$ = 80$^{\circ}$ and with $\theta_0$ = 135$^{\circ}$ and $\phi_0$ = 268$^{\circ}$, respectively.

Next, we compare the observed velocity structures of the warped structures with the model free-falling motions. Figure~\ref{img_CMU_fitting} (b and c) shows the P-V diagrams of the C$^{18}$O emission along the parabolic trajectories. These velocity structures are different from those of the outflows associated with Per-emb-8 (Fig.~\ref{img_PV_Per8_outflow}). 
The model velocity of the infalling flow in the $r$, $\theta$ and $\phi$ directions are derived as

\begin{equation}
    V_r = -\sqrt{\frac{GM_*}{r}}\cdot\sqrt{1 + \frac{\cos\theta}{\cos\theta_0}},
\end{equation}
\begin{equation}
    V_\theta = \sqrt{\frac{GM_*}{r}}\cdot\frac{\cos\theta_0 - \cos\theta}{\sin\theta}\cdot\sqrt{1 - \frac{\cos\theta}{\cos\theta_0}},
\end{equation}
\begin{equation}
    V_\phi = \sqrt{\frac{GM_*}{r}}\cdot\frac{\sin\theta_0}{\sin\theta}\cdot\sqrt{1 - \frac{\cos\theta}{\cos\theta_0}},
\end{equation}
where $\theta_0$ = 49$^{\circ}$ and $\phi_0$ = 80$^{\circ}$ are adopted for the redshifted trajectory, and $\theta_0$ = 135$^{\circ}$ and $\phi_0$ = 268$^{\circ}$ are adopted for the blueshifted trajectory. Fig.~\ref{img_CMU_fitting} shows that the model velocities can approximately explain the observed velocity structures. Therefore, the comparison between the observations and our models suggests that the infalling material toward the disk can account for the warped structures.
The mass of the arm-like structures is estimated to be $3.8\times10^{-3}$-- $5.2\times10^{-3}$ $M_{\odot}$ from the total integrated C$^{18}$O flux (12.7 Jy km s$^{-1}$), under the assumption of local thermodynamic equilibrium (LTE), the C$^{18}$O abundance of 1.8$\times10^{-7}$, and an excitation temperature of 20--30 K. Based on the infalling velocities of $\sim$1 km s$^{-1}$ and the lengths of the flows of $\sim$1000 au, the mass infalling rate is estimated to be $8\times 10^{-7}$-- $1\times 10^{-6}$ $M_{\odot}$ yr$^{-1}$.

\section{Discussion} \label{sec:discussion}
We observed the misaligned inner and outer disks in the 1.3 mm continuum emission and the multiple outflows around Per-emb-8 in the $^{12}$ CO emission. In addition, the C$^{18}$O and continuum emission shows the arm-like infalling structures around the disk of Per-emb-8. We also observed the misaligned blue- and redshifted outflow lobes associated with the binary, Per-emb-55. In this section, we will discuss the possible origins of these structures.

\subsection{Misaligned Inner and Outer Disks around Per-emb-8} \label{subsec: misaligned disks}
The 1.3 mm continuum data suggest that Per-emb-8 has a warped protostellar disk where the outer disk has a radius of $\sim$300 au and a PA of $\sim$59$^{\circ}$ while the misaligned inner disk has a radius $\sim$40 au and a PA of $\sim$103$^{\circ}$. Per-emb-8 can be an unresolved binary with the outer disk as the circumbinary disk and the inner component as the unresolved circumstellar disks (also see Sec.~\ref{subsec: multiple outflow Per-emb-8}), although the 8 mm continuum emission in Per-emb-8 only has a single peak in the VLA observations at an angular resolution of $0\farcs065$ (15 au). Misalignment of the circumbinary and circumstellar disks has been observed in several proto-binary systems \citep{2019Sci...366...90A,2020ApJ...897...59M,2016ApJ...830L..16B}, which can be due to the orbital dynamics of two protostars. Observations at even higher resolutions and sensitivity are required to examine this possibility.

If Per-emb-8 is a single protostar, the misalignment between its inner and outer disk can be generated by tidal interaction with flyby or its companion (if Per-emb-8 and 55 are gravitationally bound). Such warped disks with their inner and outer disks misaligned are often seen in theoretical simulations of tidal interaction between stars \citep{2020MNRAS.491.4108N, 2019MNRAS.483.4114C}. If Per-emb-8 has interacted with Per-emb-55 or a flyby, its disk may be warped. However, this warp feature only persists for a few thousand years in the simulations \citep{2020MNRAS.491.4108N}, so the presence of the warped disk around Per-emb-8 is less likely to be attributed to perturbation by the companion or a flyby, as no nearby stars within 1000 au are observed (also see Sec.~\ref{subsec: warped structure}).

The other possibility is that infalling material with an angular momentum of a different direction is accreted onto the disk of Per-emb-8 and causes the outer disk to be misaligned with the inner disk. 
This kind of phenomenon has been found in hydrodynamic simulations of star formation in a turbulent environment having inhomogeneous distribution of angular momentum  
\citep{2018MNRAS.475.5618B...Bate,2021A&A...656A.161K}. 
Furthermore, in these simulations, warped disks are often connected with arc or spiral structures of lengths of hundreds of au. 
Observations have suggested that multiple systems often form in turbulent environment \citep{2022ApJ...931..158L}. The warped disk around Per-emb-8 is connected with arm-like structures, and these arm-like structures are likely infalling material and but not formed by the tidal interaction with its companion (Sec.~\ref{subsec: warped structure}). Thus, this infall scenario is more likely to explain the presence of the warped disk around Per-emb-8.

%====================================================================================================
\subsection{Multiple Outflows Associated With Per-emb-8} \label{subsec: multiple outflow Per-emb-8}

Two $^{12}$CO bipolar outflows are associated with Per-emb-8 which has a warped disk. 
The north-south bipolar outflow, which is previously identified with SMA and also seen in the ALMA data, is perpendicular to the major axis of the inner disk. The northwest-southeast bipolar outflow identified in this work is perpendicular to the major axis of the outer disk and also perpendicular to the protostellar envelope observed in the CS emission in \cite{2023A&A...678A.124A}. 

The VANDAN observation at a resolution of $\sim$15 au suggests that Per-emb-8 is a single protostar \citep{2016ApJ...818...73T}. 
 Multiple outflows launched by a single protostar have also been observed in the low mass Class 0 protostar IRAS 15398-3359 \citep{2021ApJ...910...11O...Okoda}. In IRAS 15398-3359, the ALMA observation at a resolution of 0$\farcs$04 ($\sim$6 au) reveals a single peak in the dust continuum emission \citep{2023ApJ...958...60T}, while multiple outflows are detected in molecular-line emissions \citep{2021ApJ...910...11O...Okoda}. 
 The dynamical timescales of these multiple outflows are different by almost an order of magnitude 
 \citep{2014ApJ...795..152O,2016A&A...587A.145B,2021ApJ...910...11O...Okoda}. 
 \cite{2021ApJ...910...11O...Okoda} suggest that the secondary outflow in IRAS 15398-3359 is caused by a past reorientation of the outflow launching. 
In Per-emb-8, the base of both bipolar outflows is observed to be connected with the protostellar disk, given our angular resolution (Fig.~\ref{img_12CO_mom0}). The dynamical timescales of both bipolar outflows are similar, which are estimated to be 2000--3000 years from the observed lengths and velocities of the outflows. Thus, the presence of multiple outflows associated with Per-emb-8 is unlikely due to the change in the outflow launching direction with time.

Theoretical simulation shows that a bipolar outflow can be launched from a magnetized circumstellar disk via magnetocentrifugal force \citep{ 2014ApJ...796L..17M...Machida}. Such outflows launching from disks have been observed with ALMA \citep[e.g.,][]{2016Natur.540..406B, 2023ApJ...945...63H}. 
The protostellar disk around Per-emb-8 is warped. It is possible that the outflows are launched from both the misaligned inner and outer disks via magnetocentrifugal force \citep{2000prpl.conf..759K,2008ApJ...676.1088M}, and thus two misaligned bipolar outflows along the normal axes of the inner and outer disks are observed. 

Nevertheless, we cannot rule out that Per-emb-8 is a binary system with a projected separation less than 15 au, which could not be resolved with the previous VANDAN observation, or there is a low mass companion with faint dust continuum emission inside the inner disk that could not be detected by the previous VLA observation \citep{2016ApJ...818...73T}.
Binary/multiple systems are often associated with multiple outflows \citep{2020ApJ...897...59M,2015A&A...584A.126S...Santangelo, 2022ApJ...927...54O,2006ApJ...653.1358K...Kwon,2015ApJ...798...61T,2019ApJ...870...81T}.
The disk observed in the continuum emission in Per-emb-8 is better represented with the two misaligned components  (Fig.~\ref{img_Per8_cont_fitting}). The outer, more extended component could be a circumbinary disk, and the inner component could be unresolved circumstellar disks. In this case, two bipolar outflows may be launched from each of the unresolved circumstellar disks, although outflows launched from each companion in a close binary system tend to be along a similar direction \citep{2021ApJ...912...34H, 2005ApJ...630..976C}. 
The other possibility is that the north-south bipolar outflow is from one of the companions, and the northwest-southeast bipolar outflow is from the circumbinary disk. As shown in simulations, outflows can be also launched from a circumbinary disk \citep{2009ApJ...704L..10M}. 
Further observations at higher angular resolutions to resolve the base of these outflows and the inner disk in Per-emb-8 are required to investigate the origin of its two misaligned bipolar outflows.

%====================================================================================================

\subsection{Warped Accretion Flow in Per-emb-8} \label{subsec: warped structure}

Our observational results suggest that Per-emb-8 is surrounded by a Keperian disk with a radius of $\sim$300 au, and the extended, non-Keperian warped structures are observed in the C$^{18}$O and 1.3 mm continuum emission. These warped structures likely trace the infalling materials from the east and west of Per-emb-8, and their morphologies and velocity structures can be approximately explained by parabolic free-fall motion with a conserved angular momentum (Fig.~\ref{img_CMU_fitting}). 

Theoretical simulations suggest that the formation of warped accretion flows can be related to the magnetic field and turbulence in a collapsing dense core. If the magnetic field orientation and rotational axis are misaligned in the dense core, the materials would first collapse along the magnetic field and form a flattened envelope or pseudo-disk, and the flatten structure can be twisted at inner radii where the rotational motion is more dominant and appear to be warped flows in observations \citep{2013ApJ...774...82L,2020ApJ...898..118H}. On the other hand, if the dense core is turbulent, the flattened structure typically formed perpendicular to the magnetic field can be warped and broken into smaller pieces by the turbulence, and can be identified as warped accretion flows observationally \citep{2014ApJ...793..130L, 2024MNRAS.52710131T}. 
The turbulent velocity has been estimated to be 1.4 times the sound speed (or Mach number) in the dense core of Per-emb-8 \citep{2011MNRAS.410...75C}. 
This level of turbulence is comparable to those in the theoretical simulations forming warped accretion flows in turbulent dense cores \citep{2014ApJ...793..130L, 2024MNRAS.52710131T}. This suggests that the turbulence in the dense core of Per-emb-8 is strong enough to form the warped structures.

As shown in the theoretical simulations, the turbulence and magnetic field misaligned with the rotational axis can alleviate magnetic braking in the collapsing dense core and enable the formation of a sizable protostellar disk \citep{2012A&A...543A.128J, 2012ApJ...747...21S,2013MNRAS.432.3320S}. These may lead to the formation of the large disk around Per-emb-8. 
Future observations to characterize the magnetic field structures in the dense core and accretion flows are needed to investigate the relative importance between the magnetic field and turbulence in the formation of the warped accretion flows in Per-emb-8.

Per-emb-8 is in a multiple system. Theoretical simulations show that tidal interaction with a companion or flyby can warp a disk and form spiral structures stretching out from the disk \citep{2019MNRAS.483.4114C}. In these simulations, one of the prominent spiral arms formed by the tidal interaction typically points toward the perturber, 
and these spiral structures disappear or become tightly wound after a few thousand years after the encounter. 
Per-emb-8 may experience interaction with its companion, Per-emb-55. However, the warped or arm-like structures connecting the Per-emb-8 disk do not extend toward Per-emb-55, and there are no other arm-like structures connecting the two observed in the continuum and line emission. Furthermore, if Per-emb-55 indeed rotates around Per-emb-8 with a high eccentricity, the orbital period is at least a few $\times$ 0.1 Myr, given the current projected separation of 2782 au. Even if there was a close encounter between Per-emb-8 and Per-emb-55, arm-like structures formed by their tidal interaction are expected to disappear, as shown in the simulations \citep{2019MNRAS.483.4114C}. Therefore, the warped structures around the disk of Per-emb-8 are less likely due to the tidal interaction with the companion. Nevertheless, we note that the lifetime of disk structures caused by a perturber in those smooth particle hydrodynamics simulations can be underestimated because the viscosity in the simulations can be unrealistically high. More detailed comparison with the observed morphologies and velocity structures with those in the simulations is needed to further examine this possibility.

%===================================================================================================
\subsection{Asymmetric Outflows Associated With Per-emb-55} \label{subsec: Outflow Per-emb-55}
Conic redshifted and blueshifted lobes of the outflow associated with Per-emb-55 are toward the west and south, respectively (Fig.~\ref{img_12CO_mom0}), different from typical bipolar outflows having two lobes aligned. Bipolar outflows with misaligned blueshifted and redshifted lobes have been observed in several protostellar sources, which could be caused by the outflow processing, dynamical interaction with the dense cloud, or the electromagnetic interaction \citep[e.g.,][]{2014A&A...567A..99P,2018ApJ...863...19A, 2005ApJ...630..976C, 2016ApJ...819..159C}. 
Around Per-emb-55, if there is dense ambient material located in the northeast, the bipolar outflow launched from this binary system could be deflected to have its two lobes toward the west and south. 
The ALMA data cannot reveal dense ambient material even if it is present because of interferometric filtering of large-scale structure.
Observations with shorter baselines are needed to explore this possibility. 
The other possibility is that Per-emb-55 is moving toward northeast while launching this bipolar outflow. 
The outflow velocity is approximately $\sim$5 km s$^{-1}$, as seen in the P-V diagram (Fig.~\ref{img_PV_Per55_outflow}). 
If the system is moving toward northeast at a similar velocity, the two lobes of the bipolar outflow launched from it could appear to be misaligned. Future observations to measure the proper motion of Per-emb-55 can test this hypothesis.

On the other hand, since Per-emb-55 is a binary system, it is possible that the each outflow lobe is launched by each protostar. Monopolar outflows have been observed in a few protostellar sources, although they are not often seen \citep{2017ApJ...846...16S,2004A&A...426..503W}. As shown in the theoretical calculation, monopolar outflows can form depending on the 
magnetic field structures of star+disk systems and the energy ratio of the magnetic field to turbulence in protostellar envelopes
\citep{2010MNRAS.408.2083L,2024ApJ...963...20T}. It is also possible that Per-emb-55 $A$ and $B$ both actually launch bipolar outflows, but one of the lobes is obscured by extended ambient clouds if its line-of-sight velocity is coincident with the cloud velocity. The continuum results suggest that the orientation of the disk around Per-emb-55 $A$ is along the north-south direction, while the disk around Per-emb-55 $B$ is not resolved. Thus, the redshifted outflow lobe toward the west could be launched by source $A$, and the blueshifted outflow lobe toward the south by source $B$. Association between the outflows and launching sources cannot be determined with the current resolution, and high-resolution observations are needed.

%===================================================================================================

\section{Summary} \label{sec: summary}
We have analyzed the ALMA archival data of the 1.3 mm continuum, $^{12}$CO (2-1), and C$^{18}$O (2-1) emission in a proto-multiple system which consists of a Class 0 protostar Per-emb-8 and a proto-binary Class I protostars Per-emb-55 $A$ and $B$. The aim is to probe their Keplerian disks and the gas kinematics to understand the star formation process in the proto-multiple system. Our main results are summarized below.
\begin{enumerate}
    \item 
    The 1.3 mm continuum data suggests that Per-emb-8 has a warped disk with a radius of 300 au, and Per-emb-55 $A$ has a compact disk with a radius of 60 au, while the continuum emission in Per-emb-55 $B$ is not resolved at the resolution of 0\farcs3. 
    The inner and outer parts of the Per-emb-8 disk are misaligned by $\sim 40^{\circ}$, and the inner disk has a radius of 40 au. 
   This warped disk could be formed by accreting infalling material with angular momenta of different directions, and is less likely due to tidal interaction with its companion.

    \item  
    Two bipolar outflows, one along the north-south direction and the other along the northwest-southeast direction, are identified in the $^{12}$CO emission in Per-emb-8, 
    and they are along the normal axes of the inner and outer parts of the warped disk around Per-emb-8, respectively. Thus, the two bipolar outflows could be launched from the inner and outer disks of Per-emb-8. Nevertheless, we cannot rule out that Per-emb-8 is an unresolved binary with a projected separation is less than 15 au, and the two bipolar outflows are launched by the individual companions.
   
    \item The blue- and redshifted lobes of the $^{12}$CO bipolar outflow associated with Per-emb-55 are found to be misaligned, which are toward the south and west, respectively. 
    The two lobes can be misaligned, if there is a dense ambient cloud in the northeast deflecting the outflow motions, or if Per-emb-55 is moving toward the northeast at a velocity of 5 km s$^{-1}$ similar to the outflow velocity. 
    Since monopolar outflows have been observed in a few protostellar sources as well as in numerical simulations, it is also possible that both Per-emb-55 $A$ and $B$ launch monopolar outflows, and the two outflows are misaligned.

    \item The compact C$^{18}$O emission in Per-emb-8 exhibits a clear velocity gradient along the major axis of the central continuum emission, which likely traces the Keplerian rotation. By fitting the Keplerian velocity profile, we estimate the mass of Per-emb-8 to be 1.6--2.8 $M_{\odot}$. 
    Toward Per-emb-55, the $^{12}$CO emission is detected but not the C$^{18}$O emission.  
    The $^{12}$CO emission in Per-emb-55 $A$ shows a possible signature of disk rotation. 
    By comparing the P-V diagram of the $^{12}$CO emission with Keplerian rotational profiles, the stellar mass of Per-emb-55 $A$ is estimated to be 0.4--1 $M_{\odot}$.
    We estimated the stellar mass of Per-emb-55 $B$ from the $^{12}$CO spectrum because its disk is not resolved. Assuming the velocities of the peaks in the $^{12}$CO spectrum represent the Keplerian velocities and the disk size is the same as the beam size, its stellar mass is estimated to be 0.3 $M_{\odot}$. Thus, Per-emb-8 is the primary star in this triple system.

    \item Two arm-like warped structures connecting the disk and extending toward east and west with a length $\sim$1000 au are observed in the continuum and C$^{18}$O emission, and they exhibit a clear velocity gradient along the east-west direction, different from the central disk. 
    Their morphologies and velocity structures can be explained with the CMU model of infalling motion. Thus, these structures are mostly likely the infalling material toward the disk.
    These warped accretion flows could form because of misalignment between the magnetic field and rotational axis, or the turbulence in the dense core, and they are less likely due to tidal interaction with the companions in the system. 
    
\end{enumerate}
Our study reveals complex gas distributions and kinematics in this proto-multiple system Per-emb-8 and 55. 
These structures are often seen in numerical simulations of star formation in turbulent environments and those of tidal interaction in multiple systems. 
Further observations to probe the magnetic field structures in the dense core and accretion flows, resolve the base of the outflows, and detect proper motions of the companions are essential to assess the relative importance of the magnetic fields, turbulence, and tidal force and orbital motion of the companions in the formation process of multiple systems.

\begin{acknowledgments}
We thank Jeremy L.\ Smallwood for fruitful discussions on smooth particle hydrodynamics simulations and tidal interaction. 
H.-W.Y.\ acknowledges support from National Science and Technology Council (NSTC) in Taiwan through grant NSTC 110-2628-M-001-003-MY3 and NSTC 112-2124-M-001-014 and from the Academia Sinica Career Development Award (AS-CDA-111-M03).
This paper makes use of the following ALMA data: ADS/JAO.ALMA $\#$2017.1.01078.S (PI: Dominique Segura-Cox). ALMA is a partnership of ESO (representing its member states), NSF (USA) and NINS (Japan), together with NRC (Canada), NSTC and ASIAA (Taiwan), and KASI (Republic of Korea), in cooperation with the Republic of Chile. The Joint ALMA Observatory is operated by ESO, AUI/NRAO and NAOJ.
\end{acknowledgments}

\facility{ALMA}
\software{CASA \citep{2022PASP..134k4501C}, Numpy \citep{2011CSE....13b..22V,harris2020array}, SciPy \citep{2020NatAs...4.1158P}, Astropy \citep{astropy:2013, astropy:2018, astropy:2022}, Matplotlib \citep{2007CSE.....9...90H}, SLAM \citep{2023zndo...7783868A}}

\bibliography{references}{}
\bibliographystyle{aasjournal}



\end{document}